\documentclass{ws-mpla}

\begin{document}

\markboth{Pastore, Schiavilla, and Goity}
{Electromagnetic Processes in $\chi$EFT}

\catchline{}{}{}{}{}

\title{ELECTROMAGNETIC PROCESSES IN $\chi$EFT}

\author{S.\ Pastore$^{\rm a}$, R.\ Schiavilla$^{\rm a,b}$, and J.L.\ Goity$^{\rm b,c}$}

\address{$^{\rm a}$Physics Department, Old Dominion University, Norfolk, Virginia 23529, USA\\
$^{\rm b}$Theory Center, Jefferson Laboratory, Newport News, Virginia 23606, USA\\
$^{\rm c}$Physics Department, Hampton University, Hampton, Virginia 23668, USA\\
pastore@jlab.org, schiavil@jlab.org, goity@jlab.org
}

\maketitle


\begin{abstract}
Nuclear electromagnetic currents derived in a chiral-effective-field-theory
framework including explicit nucleons, $\Delta$ isobars, and pions
up to N$^2$LO, {\it i.e.} ignoring loop corrections, are used in a study of
neutron radiative captures on protons and deuterons at thermal energies,
and of $A$=2 and 3 nuclei magnetic moments.  With the strengths of the
$\Delta$-excitation currents determined to reproduce the $n$-$p$ cross section
and isovector combination of the trinucleon magnetic moments, we find that the
cross section and photon circular polarization parameter, measured respectively
in $n$-$d$ and $\vec{n}$-$d$ processes, are significantly underpredicted by theory.
\end{abstract}

\ccode{PACS Nos.: 12.39.Fe, 13.40.-f, 25.10.+s, 25.40.Lw}

\section{Introduction}
\label{sec:sec1}

Nuclear electromagnetic currents have recently been derived in a
chiral-effective-field-theory ($\chi$EFT) framework including nucleons,
$\Delta$ isobars, and pions~\cite{Pastore08}.  Formal expressions up to
one loop have been obtained in time-ordered perturbation theory with
non-relativistic Hamiltonians constructed from the chiral Lagrangian
formulation of Refs.~\cite{Weinberg90,vanKolck94,Epelbaum98}.  Thus, the study
in Ref.~\cite{Pastore08} is similar to the work of Park {\it et al.}~\cite{Park96},
albeit it uses a different formalism. 

The present talk is a much abridged summary of Ref.~\cite{Pastore08}.
The currents up to next-to-next-to-leading order (N$^2$LO), that
is ignoring loop corrections which enter at N$^3$LO, are used to
calculate the magnetic moments of $A$=2 and 3 nuclei, and the thermal
neutron radiative captures on protons and deuterons.  Realistic
two- and three-nucleon (for $A$=3) potentials are used to generate
the bound and continuum wave functions.  To have an estimate
of the model dependence arising from short-range phenomena, the
variation of the predictions is studied as function of the cutoff
parameter, which is used to regularize the two-body operators, as well
as function of the input potentials---either the Argonne $v_{18}$
(AV18)~\cite{Wiringa95} or CD-Bonn (CDB)~\cite{Machleidt01} in combination
with respectively the Urbana IX~\cite{Pudliner97} or Urbana
IX$^*$~\cite{Viviani07}---used to generate the wave functions
(the AV18 and CDB have rather different short-range behaviors).

We find that the N$^2$LO calculations do not provide a satisfactory
description of the experimental data, particularly for the $^2$H($n,\gamma$)$^3$H
process.  It remains an interesting question whether N$^3$LO corrections
will resolve the present discrepancies between theory and experiment.

\section{Currents up to N$^2$LO}
\label{sec:tree}

\begin{figure}[t]
\centerline{\psfig{file=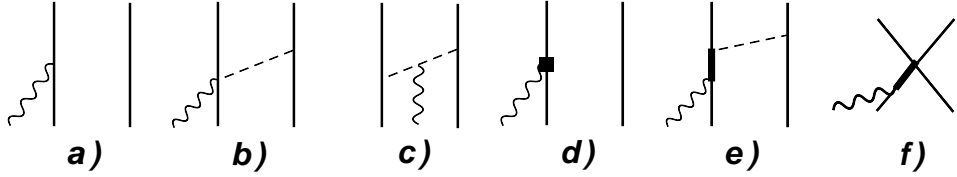,width=5.0in}}
\caption{Diagrams illustrating LO [panel a)], NLO [panels b)-c)],
and N$^2$LO [panels d)-f)] currents.  Nucleons, $\Delta$ isobars, pions, and
photons are denoted by solid, thick-solid, dashed, and wavy lines,
respectively.}
\label{fig:fig1}
\end{figure}

The currents up to N$^2$LO are illustrated by the diagrams in
Fig.~\ref{fig:fig1}, where we show only one of the possible time orderings.
The LO and NLO currents, panels a) and b)-c), are well known~\cite{Park96}
and will not be given here.  At N$^2$LO there is a contribution
originating from $(Q/M)^2$ corrections to the LO (one-body)
current ($Q$ is the low momentum scale and $M \simeq 1$ GeV is the typical hadronic
mass scale); it reads~\cite{Pastore08}:
\begin{eqnarray}
 {\bf j}^{\rm N^2LO}_{\rm RC}=&-&\frac{e}{8 \, m_N^3}
 e_{N,1}\, \Bigg[
2\, \left( K_1^2 +q^2/4 \right)
 \left( 2\, {\bf K}_1+i\, {\bm \sigma}_1\times {\bf q } \right)
+ {\bf K}_1\cdot {\bf q}\,
 \left({\bf q} +2i\, {\bm \sigma}_1\times {\bf K }_1 \right)\Bigg]
 \nonumber \\
 &-& \frac{i\,e}{8 \, m_N^3}
 \kappa_{N,1}\, \Bigg[ {\bf K}_1\cdot {\bf q}\,
 \left( 4\, {\bm \sigma}_1\times {\bf K}_1 -i\, {\bf q}\right)
 - \left(  2\, i\, {\bf K}_1 -{\bm \sigma}_1\times {\bf q} \right)\, q^2/2 \nonumber \\
 && \qquad\qquad \qquad +2\, \left({\bf K}_1\times {\bf q}\right) \, {\bm \sigma}_1\cdot {\bf K}_1
 \Bigg] + 1 \rightleftharpoons 2  \ ,
\label{eq:j1rc}
\end{eqnarray}
where the momenta ${\bf k}_i$ and ${\bf K}_i$ are defined as
${\bf k}_i={\bf p}_i^\prime-{\bf p}_i$ and ${\bf K}_i=({\bf p}_i^\prime+{\bf p}_i)/2$,
${\bf q}$ is the photon momentum, and 
\begin{equation}
e_{N,i}=(1+\tau_{i,z})/2 \ ,\qquad \kappa_{N,i}=(\kappa_S+\kappa_V \tau_{i,z})/2 \ ,
\qquad \mu_{N,i}=e_{N,i}+\mu_{N,i} \ ,
\end{equation}
with $\kappa_S$=--0.12 n.m and $\kappa_V$=3.706 n.m..

The N$^2$LO contributions, represented by diagrams e)-f) and
involving $\Delta$-isobar excitation, read in the static limit~\cite{Pastore08}:
\begin{eqnarray}
 {\bf j}^{\rm N^2LO}_{\Delta}&=& i\,\frac{ e\, C_\Delta}{9 \, m_N\, \Delta}
 \frac{{\bm \sigma}_2 \cdot {\bf k}_2}{k_2^2+m_\pi^2}
 \Big[ 4\, \tau_{2,z}\, {\bf k}_2-({\bm \tau}_1\times {\bm \tau}_2)_z
 \,{\bm \sigma}_1 \times{\bf k}_2\Big]
 \times {\bf q} + 1 \rightleftharpoons 2\ , \label{eq:jd}\\
 {\bf j}^{\rm N^2LO}_{\Delta_c}&=& -i\,\frac{ e\, C_{\Delta_c}}{9 \, m_N\, \Delta}
  \Big[ 4\, \tau_{2,z}\, {\bm \sigma}_2-
 ({\bm \tau}_1\times {\bm \tau}_2)_z\,
 {\bm \sigma}_1 \times{\bm \sigma}_2 \Big]\times {\bf q} + 1 \rightleftharpoons 2\ ,
\label{eq:jdc}
\end{eqnarray}
where $g_A$ and $F_\pi$ are the nucleon axial coupling
and pion decay constants, $C_\Delta=\mu^*\, g_A\, h_A/F_\pi^2$
and $C_{\Delta_c}=\mu^*\, D_T$, $h_A$ and $\mu^*$
are the $N$-$\Delta$ transition axial coupling constant and
magnetic moment, and $\Delta$ denotes the mass difference $m_\Delta -m_N$.

The configuration-space versions of the NLO and N$^2$LO operators
have $1/r^2$ and $1/r^3$ singularities ($r$ is the interparticle
separation), which need to be regularized in order to avoid divergencies
in the matrix elements of these operators between nuclear wave functions.
We adopt a simple regularization procedure, {\it i.e.}~a momentum-space cutoff.
While its precise functional form is arbitrary, the choice made here of a
Gaussian cutoff function, $C_\Lambda(p) = {\rm e}^{-(p/\Lambda)^2}$, with
the parameter $\Lambda \leq M$, is merely dictated by convenience,
since it leads to analytical expressions for the Fourier transforms~\cite{Pastore08}.
It is expected that this arbitrariness be of little relevance, since the
dependence of theoretical predictions on variations in the cutoff is (or should be,
see next section) largely removed by a renormalization of the theory free
parameters, which are fixed by reproducing a given set of observables.

\section{Results and Conclusions}

At N$^2$LO, the only isoscalar terms are from the (one-body) LO and
N$^2$LO-RC operators, which are independent of the cutoff $\Lambda$.
\begin{table}[bth]
\tbl{Contributions in n.m.~to the deuteron (columns labeled AV18 and CDB)
and $^3$He/$^3$H isoscalar combination (columns labeled AV18/UIX and
CDB/UIX$^*$) magnetic moments.  The experimental values are 0.8574 n.m. and
0.4257 n.m., respectively.}
{\begin{tabular}{@{}ccccc@{}} \toprule
         & AV18 & CDB & AV18/UIX & CDB/UIX$^*$ \\
\colrule
LO          &0.8469 &0.8521  &  0.4104 &  0.4183  \\
N$^2$LO-RC  &--0.0082 &--0.0080 &--0.0045  &  --0.0052 \\
\botrule
\end{tabular}}
\label{tb:tab1}
\end{table}
In Tables~\ref{tb:tab1} we list their contributions to the deuteron
magnetic moment and isoscalar combination of the $^3$He and $^3$H
magnetic moments.  The N$^2$LO-RC correction is (in magnitude) about
1\% of the LO contribution but of opposite sign, so that its
inclusion increases the difference between the measured and calculated
values.  As a result the experimental deuteron and trinucleon isoscalar
magnetic moments are underpredicted by theory at the (1.6--2.1)\% and
(3.0--4.7)\% levels, respectively, depending on whether the CDB and
CDB/UIX$^*$ or AV18 and AV18/UIX combinations are adopted in the $A$=2
and $A$=3 calculations.  We note that a recent calculation of these same
observables~\cite{Song07}, based on variational Monte Carlo (VMC) wave
functions corresponding to the AV18/UIX Hamiltonian, finds the magnitude
of the N$^2$LO-RC correction somewhat smaller in $A$=2 (--0.0069 n.m.)
and significantly larger in $A$=3 (--0.012 n.m.) than obtained here.
However, the expression for the magnetic dipole operator is different
from that resulting from Eq.~(\ref{eq:j1rc}), and the VMC wave functions are
less accurate than the hyperspherical harmonics (HH) wave functions~\cite{Kievsky08}
used in this work.

In the isovector sector, the NLO current involves the combination $g_A/F_\pi$,
for which we adopt the value $(m_\pi\, g_A/F_\pi)^2/(4\pi)$=0.075 as inferred
from an analysis of nucleon-nucleon elastic scattering data~\cite{Stoks94}.
In the N$^2$LO currents, the parameters $C_\Delta$ and $C_{\Delta_c}$ are determined
by reproducing the $n$-$p$ radiative capture cross section and $^3$He/$^3$H isovector
magnetic moment, respectively.  We note that the $\Delta_c$ current in Eq.~(\ref{eq:jdc})
gives no contribution in the $n$-$p$ capture.  It contributes in three-body matrix
elements only because in the configuration-space version of this operator the
$\delta$-function is replaced by a finite width Gaussian~\cite{Pastore08}.  It is
for this reason that one can interpret the contributions resulting from the $\Delta_c$
current as representing corrections beyond N$^2$LO.

Results for the isovector combination $\mu_V$ of the trinucleon magnetic moments
(without inclusion of the $\Delta_c$ current contribution) are presented
in Table~\ref{tb:tab5}.  The NLO contribution calculated
\begin{table}[bth]
\tbl{Contributions in units of n.m.~to the isovector combination of the trinucleon
magnetic moments, obtained with the AV18/UIX and CDB/UIX$^*$ Hamiltonian models and
cutoff values in the range 500--800 MeV.  The LO and N$^2$LO-RC contributions are
cutoff independent. The experimental value is --2.553 n.m..}
{\begin{tabular}{@{}ccccccc@{}} \toprule
&  \multicolumn{3}{c}{AV18/UIX} & \multicolumn{3}{c}{CDB/UIX$^*$} \\
\hline
 $\Lambda$ (MeV)               & 500 & 600 & 800 & 500  & 600 & 800 \\
\hline
LO               &   --2.159 & --2.159     &  --2.159  & --2.180   & --2.180 & --2.180  \\
NLO              &   --0.156 & --0.197     &  --0.238  & --0.113   & --0.156 & --0.200  \\
N$^2$LO-RC       &    +0.029 &  +0.029     &   +0.029  &  +0.024   &  +0.024 &  +0.024  \\
N$^2$LO-$\Delta$ &   --0.258 & --0.253     &  --0.250  & --0.205   & --0.202 & --0.200  \\
\hline
Sum              &   --2.544 & --2.580    &   --2.618  & --2.474   & --2.514 & --2.556  \\
\botrule
\end{tabular}}
\label{tb:tab5}
\end{table}
in Ref.~\cite{Song07} with VMC wave functions and a cutoff of 600 MeV
is --0.205 n.m., which is 4\% larger than obtained here.  Of course,
the parameter $C_{\Delta_c}$ is adjusted to reproduce, as function
of $\Lambda$, the $\mu_V$ experimental value.

Predictions for the cross section $\sigma_T$ and photon circular
polarization parameter $R_c$ measured in the reaction $^2$H($n,\gamma$)$^3$H
(with unpolarized and polarized neutrons, respectively) are presented in
Table~\ref{tb:tab9}.  At thermal energies this process proceeds through S-wave
capture predominantly via magnetic dipole transitions from the initial doublet
$J$=1/2 and quartet $J$=3/2 $n$-$d$ scattering states.  In addition, there is a
small contribution due to an electric quadrupole transition from the initial quartet
state~\cite{Viviani96}.
\begin{table}[bth]
\tbl{Cumulative contributions to the cross section $\sigma_T$ (in mb) and photon
polarization parameter $R_c$ of the reaction $^2$H($n,\gamma$)$^3$H at thermal
energies, obtained with the AV18/UIX Hamiltonian model and cutoff values in
the range 500-800 MeV.  The experimental values for $\sigma_T$ and $R_c$ are
respectively ($0.508 \pm 0.015$) mb from Ref.~\protect\cite{Jurney82}
and $-0.42 \pm0.03$. from Ref.~\protect\cite{Konijnenberg88}.}
{\begin{tabular}{@{}ccccccc@{}} \toprule
 &  \multicolumn{3}{c}{$\sigma_T$} & \multicolumn{3}{c}{$R_c$}  \\
\hline
 $\Lambda$ (MeV)   & 500 & 600 & 800 & 500  & 600 & 800  \\
\hline
LO                             & 0.229 & 0.229 & 0.229 & --0.060 & --0.060 & --0.060 \\
LO+NLO                         & 0.272 & 0.260 & 0.243 & --0.218 & --0.182 & --0.123 \\
LO$+\cdots+$N$^2$LO-RC         & 0.252 & 0.241 & 0.226 & --0.152 & --0.109 & --0.041 \\
LO$+\cdots+$N$^2$LO-$\Delta$   & 0.438 & 0.416 & 0.389 & --0.432 & --0.418 & --0.397 \\
LO$+\cdots+$N$^2$LO-$\Delta_c$ & 0.450 & 0.382 & 0.315 & --0.437 & --0.398 & --0.331 \\
\hline
\end{tabular}}
\label{tb:tab9}
\end{table}
At N$^2$LO the cross section is underpredicted by theory by (11--38)\% as
the cutoff is increased from 500 MeV to 800 MeV.  This rather drastic cutoff
dependence is mostly due to the contribution of the N$^2$LO-$\Delta_c$ current.
Indeed removing it leads to a much weaker variation of the cross section---roughly
$\pm 5$\% about the value obtained with $\Lambda=600$ MeV (next to last
row of Table~\ref{tb:tab9}).  It will be interesting to see to what extent, if any,
loop corrections at N$^3$LO will improve the present predictions, and in
particular reduce the cutoff dependence.

The photon polarization parameter is very sensitive to contributions of
NLO and N$^2$LO currents, which produce more than a sixfold increase,
in absolute value, of the LO result, and bring it into much closer agreement
with the measured value.  All results listed in Table~\ref{tb:tab9} for $R_c$
(and $\sigma_T$) include the small $e_{44}$ RME, although it only has a significant
effect for the LO prediction ($R_c$=--0.060 versus --0.072 depending
on whether $e_{44}$ is retained or not).


We conclude by summarizing our results.  Up to N$^2$LO, the only isoscalar terms
are those generated in a non-relativistic expansion of the one-body current, and
provide a (cutoff-independent) 1\% correction---relative to LO---to the deuteron
and isoscalar combination of the trinucleon magnetic moments.  This correction is
of opposite sign to the LO contribution, and therefore increases the underprediction
of the corresponding experimental values from $(0.9\pm 0.3)$\% for the deuteron and
$(2.7 \pm 0.9)$\% for the trinucleons at LO to, respectively, $(1.9 \pm 0.3)$\% and
$(3.8 \pm 0.8)$\% at N$^2$LO.  The spread reflects differences in the short-range
behavior of the AV18 and CDB potentials, in particular the weaker tensor components of
the latter relative to the former in this range.

At NLO, isovector terms arise from the pion seagull and in-flight contributions,
while at N$^2$LO, in addition to the relativistic corrections mentioned above,
isovector terms due to $\Delta$-isobar excitation are also obtained.  The parameters
$C_\Delta$ and $C_{\Delta_c}$ of the N$^2$LO two-body $\Delta$-excitation currents
have been determined, as functions of the cutoff $\Lambda$ and for the Hamiltonian
model of interest, by reproducing the cross section for the $n$-$p$ radiative capture
at thermal neutron energies and the isovector combination of the trinucleon magnetic
moments.  This current has then been used to make predictions---with the AV18/UIX model
only, since HH continuum wave functions are not yet available for the CDB/UIX$^*$ model---for
the cross section $\sigma_T$ and photon circular polarization parameter $R_c$ measured
in the capture of, respectively, unpolarized and polarized neutrons by deuterons.  The
experimental $\sigma_T$ ($|R_c|$) is found to be underestimated by 11\%
(overestimated by 4\%) for $\Lambda$=500 MeV and 38\% (underestimated by 21\%) for
$\Lambda$=800 MeV.  

The results display a significant cutoff dependence, particularly so for the
N$^2$LO contributions associated with $\Delta$ isobar degrees of freedom.
Indeed these contributions are much larger than those at NLO.  This is partly
due to the fact that the two NLO (pion seagull and in-flight) terms interfere
destructively.  For example, the seagull (in-flight) contributions to doublet $m_{22}$
and quartet $m_{44}$ M1 matrix elements, in units of fm$^{3/2}$ and
for $\Lambda$=500 MeV, are respectively
--9.1 (+6.5) and --0.8 (+0.6).  As a result $\sigma_T=0.425$ mb and $R_c=-0.425$
at LO+NLO (seagull only), which should be compared to $\sigma_T=0.272$ mb and
$R_c=-0.218$ at LO+NLO (seagull+in-flight) from the second row of Table~\ref{tb:tab9}.
The relatively large $\Delta$-excitation contributions also point to the
need for including loop corrections at N$^3$LO, which these N$^2$LO
currents, because of the procedure adopted here to determine their strength,
are implicitly making up for.

The next stage in the research program we have undertaken is to incorporate
the N$^3$LO operators derived in Ref.~\cite{Pastore08} into the calculations of the captures and
magnetic moments involving light nuclei (with mass number $A \leq 8$), and indeed
to extend these calculations to also include $p$-$d$ capture at energies up to a
few MeV's, and possibly four-nucleon processes, in particular $^3$He($n,\gamma$)$^4$He
at thermal energies.  Of course, at N$^3$LO three-body currents also occur, and
will need to be derived.  Work along these lines is being pursued vigorously.

\section*{Acknowledgments}

We would like to thank E.\ Epelbaum, L.\ Girlanda, A.\ Kievsky,
L.E.\ Marcucci, and M.\ Viviani for discussions.  The work of R.S.\ is supported by
the U.S.~Department of Energy, Office of Nuclear Physics, under contract DE-AC05-06OR23177,
while that of J.L.G.\ by NSF grant PHY-0555559.  The calculations were made possible
by grants of computing time from the National Energy Research Supercomputer Center.


\begin{thebibliography}{0}
%
\bibitem{Pastore08}
S.\ Pastore, R.\ Schiavilla, and J.\ Goity, in preparation.
%
\bibitem{Weinberg90}
S.\ Weinberg,
Phys.\ Lett.\ {\bf B251}, 288 (1990);
Nucl.\ Phys.\ {\bf B363}, 3 (1991);
Phys.\ Lett.\ {\bf B295}, 114 (1992).
%
\bibitem{vanKolck94}
U.\ van Kolck,
Phys.\ Rev.\ C {\bf 49}, 2932 (1994);
C.\ Ord\'onez, L.\ Ray, and U.\ van Kolck,
Phys.\ Rev.\ C {\bf 53}, 2086 (1996).
%
\bibitem{Epelbaum98}
E.\ Epelbaum, W.\ Gl\"ockle, and U.-G. Meissner,
Nucl.\ Phys.\ {\bf A637}, 107 (1998);
Nucl.\ Phys.\ {\bf A747}, 362 (2005).
%
\bibitem{Park96}
T.-S.\ Park, D.-P.\ Min, and M.\ Rho,
Nucl.\ Phys.\ {\bf A596}, 515 (1996).
%
\bibitem{Wiringa95}
R.B.\ Wiringa, V.G.J.\ Stoks, and R.\ Schiavilla,
Phys. Rev.  C {\bf 51}, 38 (1995).
%
\bibitem{Machleidt01}
R.\ Machleidt,
Phys.\ Rev.\ C {\bf 63}, 024001 (2001).
%
\bibitem{Pudliner97}
B.S.\ Pudliner {\it et al.},
Phys.\ Rev.\ C {\bf 56}, 1720 (1997).
%
\bibitem{Viviani07}
M.\ Viviani {\it et al.}.
Phys.\ Rev.\ Lett.\ {\bf 99}, 112002 (2007).
%
\bibitem{Song07}
Y.-H.\ Song, R.\ Lazauskas, T.-S.\ Park, and D.-P.\ Min,
Phys.\ Lett.\ {\bf B656}, 174 (2007).
%
\bibitem{Kievsky08}
A. Kievsky, S. Rosati, M. Viviani, L.E. Marcucci, and L. Girlanda,
J. Phys. G {\bf 35}, 063101 (2008).
%
\bibitem{Stoks94}
V.G.J.\ Stoks, R.A.M.\ Klomp, C.P.F.\ Terheggen, and J.J.\ deSwart,
Phys. Rev. C {\bf 49}, 2950 (1994).
%
\bibitem{Viviani96}
M.\ Viviani, R.\ Schiavilla, and A.\ Kievsky,
Phys.\ Rev.\ C {\bf 54}, 534 (1996).
%
\bibitem{Jurney82}
E.T.\ Jurney, P.J.\ Bendt, and J.C.\ Browne,
Phys.\ Rev.\ C {\bf 25}, 2810 (1982).
%
\bibitem{Konijnenberg88}
M.W.\ Konijnenberg {\it et al.},
Phys.\ Lett.\ {\bf B205}, 215 (1988).
%
%
%
%
\end{thebibliography}
\end{document}